# A systematic Investigation of Thermoelectric Properties of Monolayers of $ZrX_2N_4$ (X = Si, Ge)

Chayan Das [a], Dibyajyoti Saikia [a], Satyajit Sahu [a]*

[a] *Department of Physics, Indian Institute of Technology Jodhpur, Jodhpur 342030, India*

## Abstract

In the past decade, it has been demonstrated that monolayers of metal dichalcogenides are well-suited for thermoelectric applications. $ZrX_2N_4$ (X = Si, Ge) is a reasonable choice for thermoelectric applications when considering a favorable value of the figure of merit in two-dimensional (2D) layered materials. In this study, we examined the thermoelectric characteristics of the two-dimensional monolayer of $ZrX_2N_4$ (where X can be either Si or Ge) using a combination of Density Functional Theory (DFT) and the Boltzmann Transport Equation (BTE). A thermoelectric figure of merit (*ZT*) of 0.90 was achieved at a temperature of 900 K for p-type $ZrGe_2N_4$, while a *ZT* of 0.83 was reported for n-type $ZrGe_2N_4$ at the same temperature. In addition, the $ZrGe_2N_4$ material exhibited a thermoelectric figure of merit (*ZT*) of around 0.7 at room temperature for the p-type. Conversely, the $ZrSi_2N_4$ exhibited a relatively lower thermoelectric figure of merit (*ZT*) at ambient temperature. At higher temperatures, the *ZT* value experiences a substantial increase, reaching 0.89 and 0.82 for p-type and n-type materials, respectively, at 900 K. Through our analysis of the electronic band structure, we have determined that $ZrSi_2N_4$ and $ZrGe_2N_4$ exhibit indirect bandgaps (BG) of 2.74 eV and 2.66 eV, respectively, as per the Heyd-Scuseria-Ernzerhof (HSE) approximation.

## 1. Introduction

The ongoing quest for materials aims to optimize the efficiency of thermoelectric generators. The majority of thermoelectric materials consist of 2D transition metal dichalcogenides (TMDC), metal alloys [1–6], and oxides [7–9] comprise most of the thermoelectric material regime. $Bi_2Te_3$ [10], PbTe [11], and SnSe [12] possess exceptional *ZT* values among them. Following the discovery of graphene, 2D transition metal dichalcogenides (TMDCs) have demonstrated significant potential for various applications, thanks to their remarkable electronic [13], optical [14], and thermal [14] properties. A 2D monolayer of MoSi2N4 was created on a centimeter scale using the chemical vapor deposition (CVD) process [15]. Consequently, we are driven to investigate the characteristics of these monolayers. These materials exhibit exceptional dynamic stability and outstanding mechanical characteristics [16]. These materials can be used to generate electricity from waste heat. Renewable energy sources are essential in the present era due to the depletion of conventional energy sources. In this work, We conducted a study on 2D monolayer of ZrSi2N4 and ZrGe2N4 using Boltzmann transport theory in association with DFT, and our findings revealed remarkable Seebeck coefficients ($S$). The Seebeck coefficient of a material quantifies the voltage produced when a temperature gradient is applied across the material. Generally, efficient thermoelectric materials should have a combination of high electrical conductivity ($\sigma$) and $S$, along with a low thermal conductivity ($k = k_e + k_l$). The two components, $k_e$, and $k_l$, indicate the respective contributions of electrons and phonons (due to lattice) to the overall thermal conductivity. These 2D monolayers exhibit exceptional mobility, remarkable stability, and favorable optoelectronic and piezoelectric characteristics [16]. These materials can be represented by N-X-N-M-N-X-N. A metal atom (M) was positioned between two nitrogen (N) atoms, and the entire system was situated between a buckled honeycomb XN layer, where X represents either Si or Ge. M represents the transition metal atom, such as Mo, W, Ti, Zr, and others. Chemical vapor deposition (CVD) can be used to produce them. Yi-Lun Hong and his coworkers successfully synthesized MoSi2N4 by chemical vapor deposition (CVD)[15]. Bohayra Mortazavi predicted mobility of 490 and 2190 $cm^2V^{-1}s^{-1}$ for p and n-type of monolayer MoSi2N4 [16]. Till now, researchers have conducted numerous commendable studies on 2D TMDC materials. Huang et al. and his group reported a very low $ZT$ value of less than 0.2, along with a high thermal conductivity of 60 W/m K for MoSe2 monolayer. SD have published a study demonstrating a remarkably high $ZT$ value of approximately 0.9 at a temperature of 600 K using a monolayer of ZrS2. Additionally, they observed a thermal conductance of 47.8 W/K. Monolayer MoS2 and WS2 have been reported to have extremely high thermal conductivities, with values of 23.15 W/mK and 72 W/mK, respectively [17]. However, monolayer HfS2 has been reported to have a relatively low thermal conductivity ($k_l$) of 2.83 W/mK, while also exhibiting a high figure of merit for thermoelectric performance ($ZT$) of 0.90 [18]. Ongoing research is focused on enhancing the thermoelectric figure of merit (*ZT*) of materials. This study rigorously examined the electrical, optical, and thermoelectric characteristics of $ZrSi_2N_4$ and $ZrGe_2N_4$ monolayers utilizing DFT (Density Functional Theory) and BTE (Boltzmann Transport Equation) methods, providing comprehensive analysis. The $ZrGe_2N_4$ monolayer exhibited a remarkable *ZT* value of 0.90 for p-type (0.83 for n-type) at a temperature of 900 K. Similarly, the $ZrSi_2N_4$ monolayer exhibited a slightly lower thermoelectric figure of merit (*ZT*) of 0.89 for p-type and 0.82 for n-type at a temperature of 900 K. Therefore, it is necessary to conduct a theoretical analysis of the thermoelectric and optical properties in order to comprehend the underlying physical and chemical characteristics that contribute to the difference in their effectiveness for thermoelectric applications. $ZrSi_2N_4$ and $ZrGe_2N_4$ exhibited significant absorption in the blue area (452 nm and 466 nm) of the visible spectrum, indicating their suitability for optoelectronic applications within the visible range.

## 2. Methodology

We conducted initial calculations using Density Functional Theory (DFT) with Projector Augmented Wave (PAW) potentials [19,20] and Perdew-Burke-Ernzerhof (PBE) functionals for the generalized gradient approximation (GGA) [21] Quantum Espresso (QE) software package. In order to prevent the interaction between two layers, we maintained a vacuum with a thickness of 30 Å between the two layers along the z-direction. The geometry optimization was conducted with a 15×15×1 k-mesh grid. The computations were performed with a wave function energy cutoff of 50 Ry and a self-consistency threshold of $10^{-9}$ Ry. The atoms were allowed to reach a force convergence threshold of $5×10^{-3}$ eV/Å after being loosened. The phonon band structure was determined by evaluating an 8×8×1 q-grid using the Density Functional Perturbation Theory (DFPT) as implemented in the QE program. The band structure and density of states (DOS) were calculated using the Heyd-Scuseria-Ernzerhof (HSE) hybrid functional [22] which was implemented in the Vienna ab-initio Simulation Package (VASP) [23]. Here Γ centered 15×15×1 k-meshes [24] were used with an electronic kinetic energy cutoff of 500 eV and energy convergence value of $10^{-6}$ eV. Thermoelectric parameters were obtained using constant scattering time approximation from BoltzTraP code [25] using the Boltzmann transport equation.

$$\sigma_{l,m} = \frac{1}{\Omega} \int \sigma_{l,m}(\varepsilon) \left[-\frac{\Delta f_\mu(T,\varepsilon)}{\Delta \varepsilon}\right] d\varepsilon \quad (4)$$

$$k_{l,m}(T,\mu) = \frac{1}{e^2 T\Omega} \int \sigma_{l,m}(\varepsilon)(\varepsilon-\mu)^2 \left[-\frac{\Delta f_\mu(T,\varepsilon)}{\Delta \varepsilon}\right] d\varepsilon \quad (5)$$

$$S_{l,m}(T,\mu) = \frac{(\sigma^{-1})_{n,l}}{eT\Omega} \int \sigma_{n,m}(\varepsilon)(\varepsilon - \mu)\left[-\frac{\Delta f_\mu(T,\varepsilon)}{\Delta \varepsilon}\right] d\varepsilon \tag{6}$$

By utilizing these equations, we have derived the transport properties. Here, $\sigma_{l,m}$, $k_{l,m}$, $S_{l,m}$, represents the electrical conductivity, thermal conductivity, and Seebeck coefficient, respectively. The symbols $e, \mu, \Omega, T$ represents the electron charge, chemical potential, unit cell volume, and temperature, respectively. The $k_l$ was evaluated using phono3py package combined with QE. In phono3py, a 2×2×1 supercells were created and evaluated with 9×9×1 k-mesh. The self-consistent calculations were then conducted using a default displacement of 0.06 Å.

## 3. Result and Discussions

### 3.1. Structural Properties and Stability

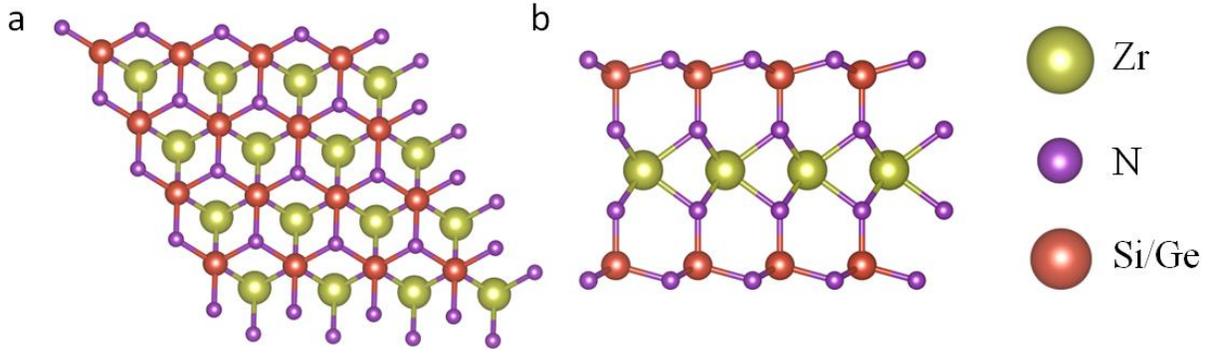

**Figure 1:** (a) A view from above and (b) a view from the side of a single layer of $ZrX_2N_4$ with a honeycomb pattern of 4x4x1 supercell.

$ZrX_2N_4$(X = Si, Ge) monolayer can be viewed as a $ZrN_2$ monolayer sandwiched between two honeycomb (SiN/GeN) layers. The three parts are stacked on top of each other. Zr atom is in the middle of a trigonal prism building block that has six Si atoms, and the $ZrN_2$ layer is bonded to (SiN/GeN) layers via vertically aligned Si–N bonds (**Figure** 1(a)). The basic unit cell of $ZrX_2N_4$ is hexagonal and has space groups of P-6m2 (No. 187) [16] (**Figure** 1(b)). We found the lattice constant for $ZrSi_2N_4$ as a = b = 3.03 Å, This is the same as what was found earlier [16]. For $ZrGe_2N_4$, the lattice constants were a = b = 3.17 Å. In Table 1, we provided the lattice constants, bond lengths, and bond angles for both structures. The bond length of $ZrGe_2N_4$ is greater than that of $ZrSi_2N_4$ due to the larger atomic radius of the Ge atom compared to the Si atom.

**Table-1:** Obtained lattice constants, bond lengths, thickness, and bond angles for both the monolayers.

| Structure | a=b (Å) | d(Zr-N) (Å) | d(X-N) (Å) | Thickness (Å) | θ(N-Zr-N) | θ(N-X-N) | θ(Zr-N-X) |
|---|---|---|---|---|---|---|---|
| ZrSi$_2$N$_4$ | 3.03 | 2.17 | 1.75 | 10.97 | 72.51 ° | 105.36 ° | 126.24 ° |
| ZrGe$_2$N$_4$ | 3.17 | 2.20 | 1.90 | 11.17 | 68.02° | 108.52 ° | 124.01 ° |

The stability of a crystalline material can be determined by evaluating its cohesive energy ($E_{ch}$) using the following formula: The equation for calculating the energy of a compound is given by the formula, $E_{ch} = \{(E_{Zr} + 2 \times E_X + 4 \times E_N) - E_{ZrX_2N_4}\}/7$. Here $E_{ZrX_2N_4}$, $E_{Zr}$, $E_X$, and $E_N$ are the energy of the monolayer of ZrX$_2$N$_4$, the energy of an individual Zr atom, individual X atom, and individual N atom, respectively. The cohesive energy per atom for the monolayers of ZrSi$_2$N$_4$ and ZrGe$_2$N$_4$ was determined to be 0.99 eV and 0.41 eV, respectively. This verifies the energetic stability of these structures. We computed the phonon dispersion curve for ZrSi$_2$N$_4$ and ZrGe$_2$N$_4$ and presented the dynamical stability of both structures along the high symmetry path of Γ-M-K-Γ, as depicted in Figure 2, As the phonon dispersion curves did not exhibit any imaginary frequencies. ZrSi$_2$N$_4$ and ZrGe$_2$N$_4$ have maximum phonon frequencies of 29.29 THz and 25.19 THz, respectively. These findings indicate that ZrGe$_2$N$_4$ has a lower frequency compared to ZrSi$_2$N$_4$. ZrGe$_2$N$_4$ exhibits some overlap between the acoustic and near-optical branches, whereas this overlap is absent in ZrSi$_2$N$_4$. The acoustic modes of ZrSi$_2$N$_4$ are primarily influenced by Zr, while the optical modes are influenced by Si and N. However, in the case of ZrGe$_2$N$_4$, the element Ge had a greater contribution to the acoustic modes, along with Zr, but a lesser contribution to the optical modes compared to ZrSi$_2$N$_4$. We conducted ab-initio molecular dynamics (MD) simulations for both structures in a micro-canonical ensemble, lasting over 4500 fs at a temperature of 900 K, as depicted in Figures 2c and 2d. This indicates that they have a strong resistance to changes in temperature and can be effectively used in actual applications at temperatures as high as 900 K. The unit cell of the ZrX$_2$N$_4$ monolayer consists of seven atoms, resulting in a total of twenty-one vibrational modes. The first three modes have lower frequencies and are classified as acoustic modes, while the remaining eighteen modes have higher frequencies and are classified as optical modes. The three vibrational modes consist of the in-plane longitudinal acoustic (LA) mode, the transverse acoustic (TA) mode, and the out-of-plane (ZA) mode.

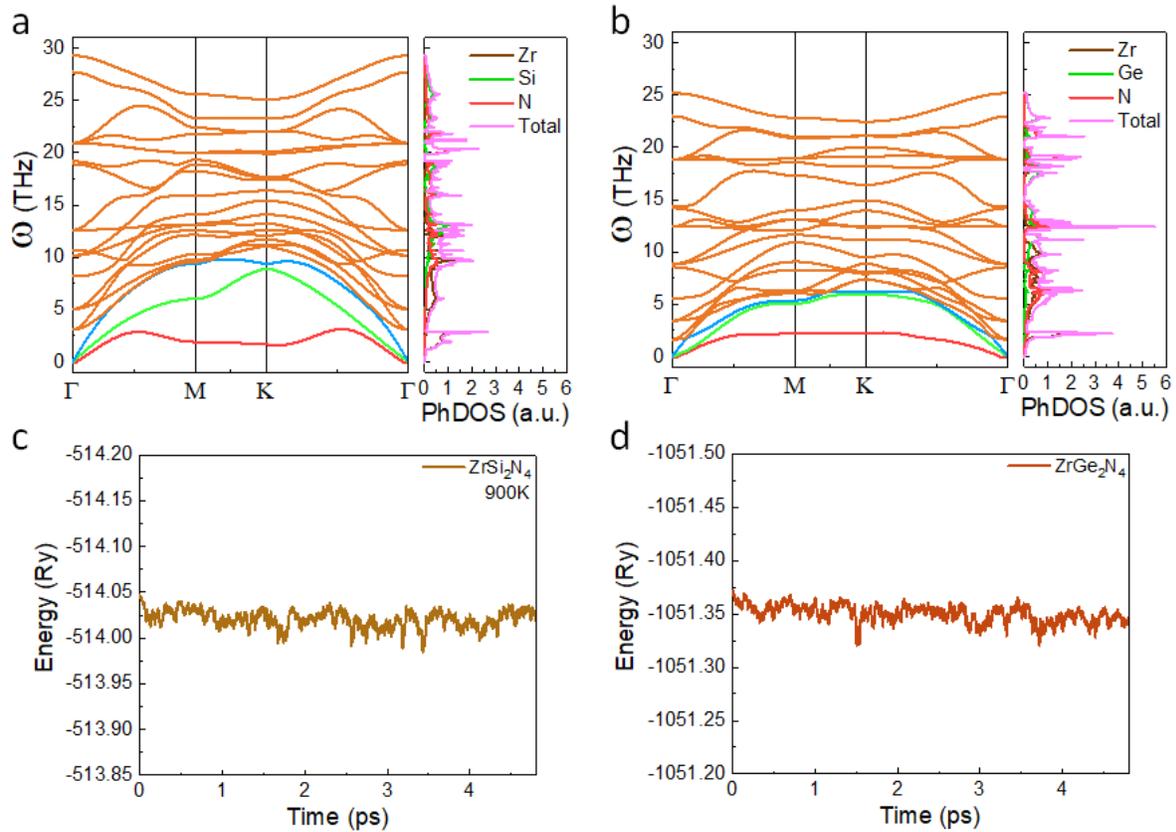

**Figure 2:** The phonon band structure and PhDOS of the monolayers of a) $ZrSi_2N_4$ and b) $ZrGe_2N4$ are presented. The bands colored in red, green, and blue correspond to the ZA, TA, and LA bands of the acoustic mode, respectively. Energy changes as a function of the time step in molecular dynamics simulations at a temperature of 900 K for c) $ZrSi_2N_4$ and d) $ZrGe_2N_4$.

### 3.2. Electronic Properties

During the investigation of the electronic band structure along the Γ-M-K-Γ path, it was shown that both $ZrSi_2N_4$ and $ZrGe_2N_4$ monolayers have an indirect bandgap (BG). This is illustrated in Figures 3a and 3b. The PBE functionals yielded band gaps of 1.63 eV and 1.57 eV, respectively. The electrical band structure of $ZrSi_2N_4$ and $ZrGe_2N_4$ was illustrated in Figures 3c and 3d. The band gap (BG) values for ZrSi$_2$N$_4$ and ZrGe$_2$N$_4$ were predicted to be 2.68 eV and 1.53 eV, respectively, using the hybrid functional (HSE06). The BG values observed for both materials were in exact accordance with the prior findings of Bohayra Mortazavi et al. [16]. The conduction band minimum (CBM) is located near the K point for both monolayers. The ZrSi$_2$N$_4$ monolayer has its valence band maximum (VBM) positioned precisely at the Γ point. However, in the case of the ZrGe$_2$N$_4$ monolayer, the VBM is located in close proximity to the Γ point, specifically between the Γ and K points. Analysis of the DOS and LDOS reveals that in ZrSi$_2$N$_4$, the valence band maximum (VBM) is predominantly influenced by the p orbitals of nitrogen and the d orbitals of the zirconium atom. On the other hand, the conduction band minimum (CBM) is principally influenced by the d orbitals of the zirconium atom. The p orbitals of the N atom provide the main contribution to the valence band maximum (VBM) in ZrGe$_2$N$_4$, while the d orbitals of the Zr atom predominantly contribute to the conduction band minimum (CBM). It has been noted that in both the monolayer, the element N has a greater impact on the valence band, while the element Zr has a greater impact on the conduction band. The p orbital of Silicon (Si) and Germanium (Ge) make about equal contributions to the Valence Band Maximum (VBM) and Conduction Band Minimum (CBM). Additionally, more precise calculations of the density of states (DOS) were demonstrated using the hybrid functional HSE06 for both structures. These calculations exhibited the same pattern as the DOS obtained using the PBE functionals.

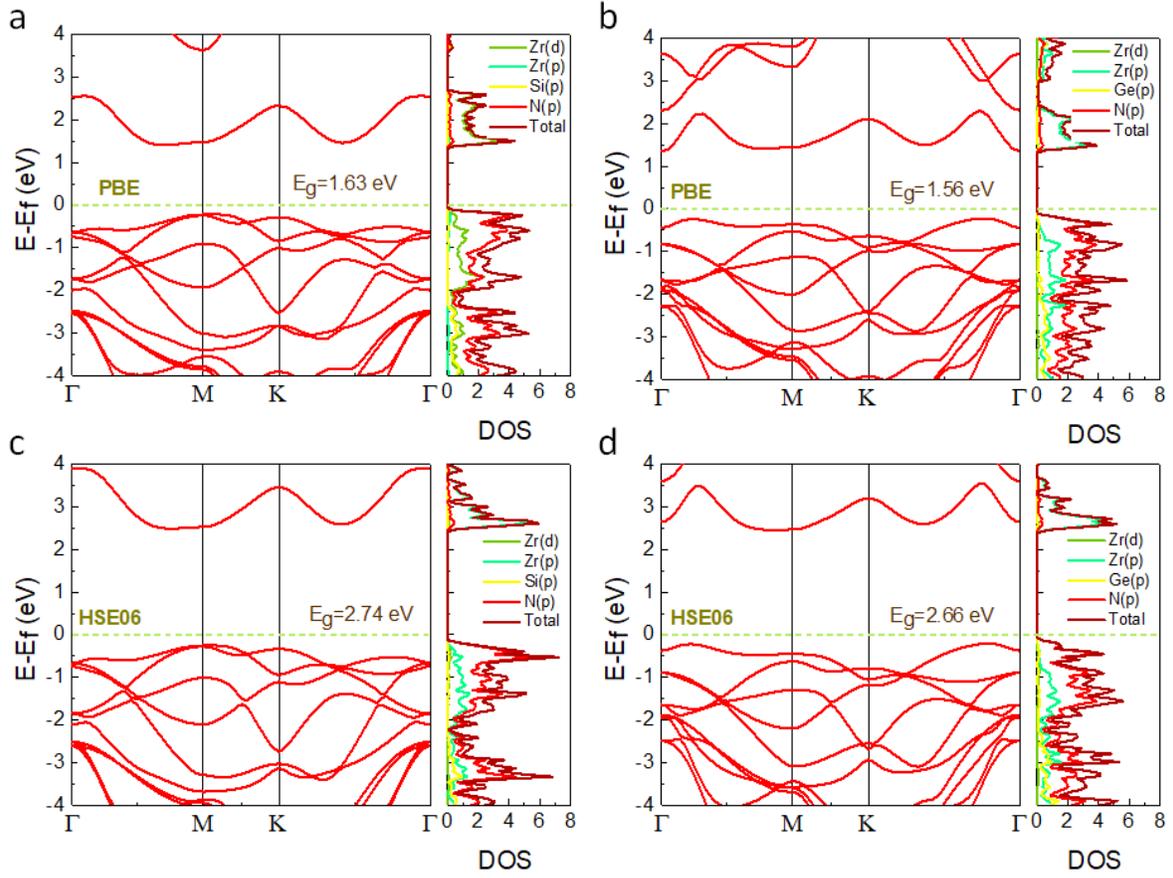

**Figure 3:** The band structure and local density of states (LDOS) of the monolayer materials (a) ZrSi$_2$N$_4$ and (b) ZrGe$_2$N$_4$ were computed using the PBE functional. Additionally, the band structure and LDOS of (c) ZrSi$_2$N$_4$ and (d) ZrGe$_2$N$_4$ were also estimated using the hybrid functional HSE06. The green dotted lines indicate the fermi level in the band gap region.

### 3.3. Carrier mobility and relaxation time

The carrier mobility of electrons and holes was determined using Bardeen and Shockley's deformation potential theory, which was applied in QE with the PBE method. The change in the band edge due to strain can be attributed to the crystal's deformation potential [26]. The mobility of 2D systems can be determined by utilizing the effective mass theorem and calculating the deformation potential, given as

$$\mu_{2D} = \frac{e\hbar^3 C_{2D}}{K_B T m_e^* m_d (E_l^2)} \tag{7}$$

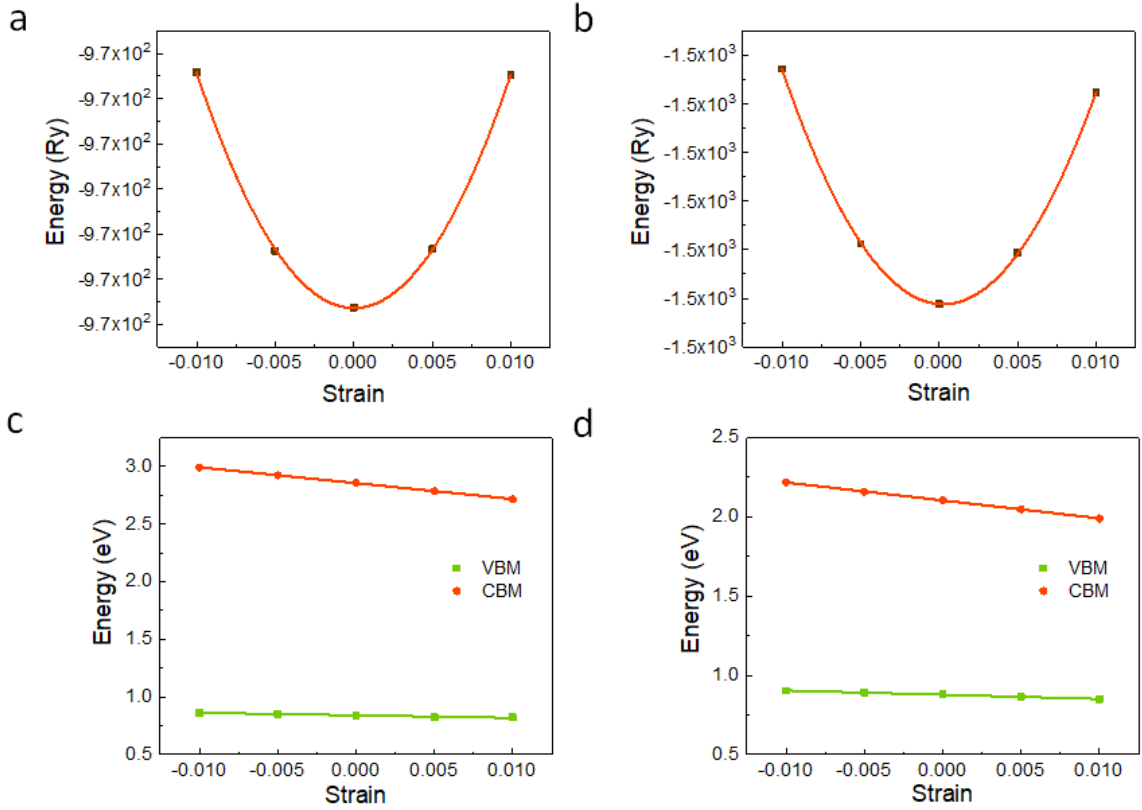

**Figure 4:** The relationship between total energy and uniaxial strain in a) $ZrSi_2N_4$ and b) $ZrGe_2N_4$. Deformation potential determination by plotting the VBM and CBM shifts against uniaxial strain in (c) $ZrSi_2N_4$ and (d) $ZrGe_2N_4$.

The elastic constant of $ZrSi_2N_4$ and $ZrGe_2N_4$ monolayers in two dimensions is denoted as $C_{2D}$ here. It can be derived from the following expression $C_{2D} = (\frac{\partial^2 E}{\partial \varepsilon^2}/A_0)$ by fitting a parabola to the strain vs. energy plot (of $ZrSi_2N_4$ (**Figure** 4a) and for $ZrGe_2N_4$ (**Figure** 4b), where $A_0$ is the area of the unstrained system of individual monolayers. In addition, T stands for the Kelvin temperature, $K_B$ is the Boltzmann constant, and $m^*(=\frac{1}{\hbar^2}\frac{d^2E}{dk^2})$ is the effective mass of various carriers and $m_d = \sqrt{m_x^* m_y^*}$. The effective mass can be determined by fitting the parabola against the band edge of VBM (for hole) and CBM (for electron). We obtained the $C_{2D}$ and hole mobility values for $ZrSi_2N_4$ were determined to be 365.22 N/m and 826.37 cm²V⁻¹s⁻¹. The $C_{2D}$ and hole mobility values of $ZrGe_2N_4$ were determined to be 306.22 N/m as 1615.02 cm²V⁻¹s⁻¹, which was nearly equal to the previously reported results [16,27–29]. The $E_I = \frac{\partial V}{\partial \varepsilon}$ is called the deformation potential, is used to describe the relationship between strain (ε) and the slope of the band edges (V). The variation of band edges of VBM and CBM for $ZrSi_2N_4$ and $ZrGe_2N_4$ is shown in **Figures 4(c) and 5(d)**. Relaxation time (τ) of individual charge carriers (for both the monolayers) can be found from the following expression $\tau = \frac{\mu m^*}{e}$. The computed parameters for both crystals are enumerated in Table 2, where $m_0$ represents the rest mass of electron. These materials exhibit significantly greater mobility compared to several other 2D materials like $MoS_2$ (72 $cm^2V^{-1}s^{-1}$) and $WS_2$ (21 $cm^2V^{-1}s^{-1}$) [30]. Their mobility is close to boron phosphorous (BP) and boron arsenide (BAs), but these materials possess very high lattice thermal conductivity of the order of $4\times10^2$ W/mK. This high thermal conductivity renders them unsuitable for thermoelectric applications [31].

**Table-2:** Calculated elastic constants, deformation potential constants, effective mass, mobility, and relaxation times of monolayer of $ZrSi_2N_4$ and $ZrGe_2N_4$ with PBE.

| Crystal | $C_{2D}$ (N/m) | $E_{I,(electron)}$ (eV) | $E_{I,(hole)}$ (eV) | $m_e^*$ /$m_0$ | $m_h^*$ /$m_0$ | $\mu_e$ ($cm^2V^{-1}s^{-1}$) | $\mu_h$ ($cm^2V^{-1}s^{-1}$) | $\tau_e$ (s) $\times 10^{-13}$ | $\tau_h$ (s) $\times 10^{-13}$ |
|---|---|---|---|---|---|---|---|---|---|
| $ZrSi_2N_4$ | 365.22 | 6.30 | 1.80 | 1.00 | 1.69 | 193.11 | 826.37 | 1.10 | 4.29 |
| $ZrGe_2N_4$ | 306.22 | 4.89 | 1.28 | 1.18 | 1.56 | 193.41 | 1615.02 | 1.30 | 16.6 |

## 3.4. Thermoelectric Properties

Figure 5a displays the fluctuation of the Seebeck coefficient (S) with respect to chemical potential (μ) at three distinct temperatures: 300 K, 600 K, and 900 K. The monolayer of $ZrSi_2N_4$ is the subject of this plot. The Seebeck coefficient (S) of the $ZrSi_2N_4$ monolayer is -2651.72 μV/K for n-type carriers (μ>0) and 2625.84 μV/K for p-type carriers (μ<0) at a temperature of 300 K. The |S| was at its most excellent at 300 K, and it diminishes as the temperature increases. Figure 5b displays the relationship between the electrical conductivity (σ/τ) scaled by the relaxation time and the parameter μ. No such change was seen in (σ/τ) with a change in temperature (T). The p-type $ZrSi_2N_4$ exhibited significantly poorer (σ/τ) compared to the n-type $ZrSi_2N_4$. Figure 5c illustrates the relationship between the relaxation time-scaled power factor (PF = $S^2\sigma/\tau$) as a function of μ for the $ZrSi_2N_4$ monolayer. The maximum power factor obtained in $ZrSi_2N_4$ monolayer was $4.12\times10^{11}$ $W/m^2Ks$ for n-type carriers ($4.19\times10^{11}$ $W/m^2Ks$ for p-type) at 900 K. **Figures** 5d, e, and f represents the variation of $S$, $\sigma/\tau$ and $S^2\sigma/\tau$ as a function of μ is shown in the ZrGe$_2$N$_4$ monolayer. The S value rapidly drops as the temperature increases in the ZrSi$_2$N$_4$ monolayer also. The maximum achieved power factor (PF) for the ZrGe$_2$N$_4$ monolayer is 1.83 ×10$^{11}$ W/m$^2$Ks, for p-type carriers, whereas for n-type carriers, the power factor attained is 4.05 ×10$^{11}$ W/m$^2$Ks at a temperature of 900 K. The efficiency of p-type doping in ZrGe$_2$N$_4$ for thermoelectric applications is significantly higher. The Seebeck coefficient (S) of ZrSi$_2$N$_4$ monolayer was discovered to be higher than that of ZrGe$_2$N$_4$ due to the direct proportionality between S and BG [32]. This can be attributed to the higher BG value of ZrSi$_2$N$_4$ compared to ZrGe$_2$N$_4$.

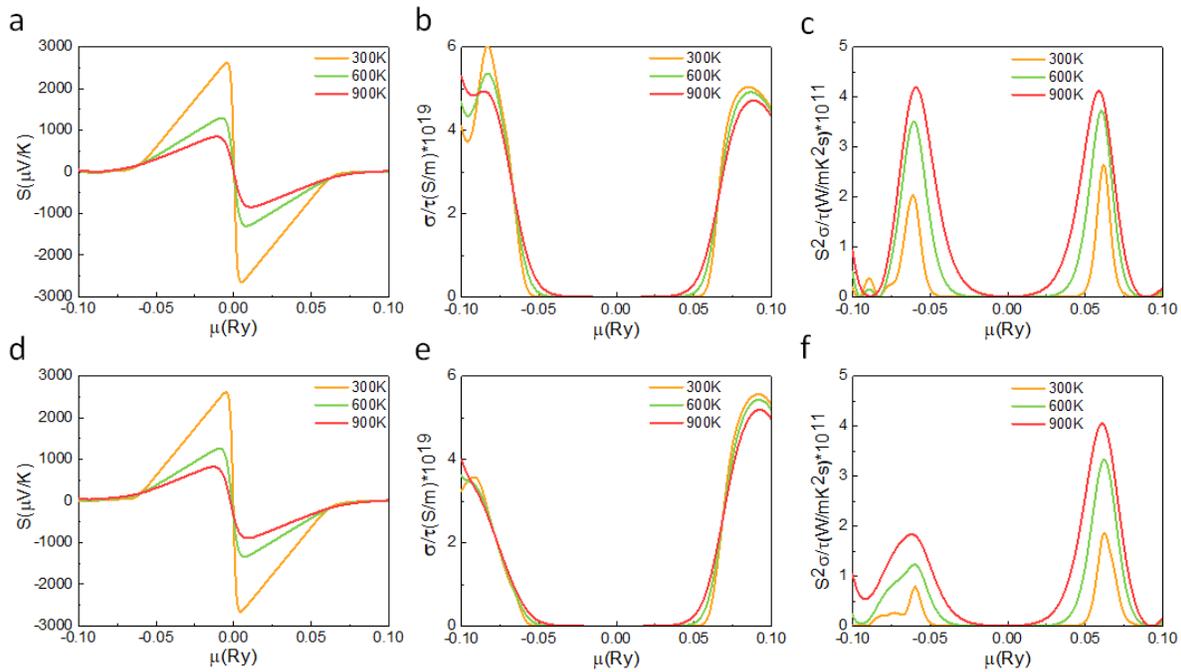

**Figure 5:** The thermoelectric characteristics, namely the Seebeck coefficients (a), electrical conductivity (b), and power factor (c) for the ZrSi$_2$N$_4$ monolayer, are represented as a function of the chemical potential μ. Similarly, the Seebeck coefficients (d), electrical conductivity (e), and power factor (f) for the ZrGe$_2$N$_4$ monolayer are also represented as a function of the chemical potential μ.

### 3.5. Lattice Thermal Conductivity ($\kappa_l$)

Figures 6a and 6b display the relationship between the temperature (T) and the fluctuation of $\kappa_l$ for monolayers of ZrSi$_2$N$_4$ and ZrGe$_2$N$_4$. The monolayer of ZrGe$_2$N$_4$ exhibited a considerably lower thermal conductivity ($\kappa_l$) of 22.08 W/m K at 300 K, in comparison to ZrSi$_2$N$_4$ (32.73 W/m

K) and even lower than the well-known 2D transition metal dichalcogenides (TMDC) like $MoS_2$ (34.5 W/(m.K)) [33], and $WS_2$ (72 W/(m.K)) [17], etc. Both structures exhibit a drop in $\kappa_l$ as temperature increases. At elevated temperatures, the $ZrGe_2N_4$ monolayer had a poorer thermal conductivity compared to the $ZrSi_2N_4$ monolayer.

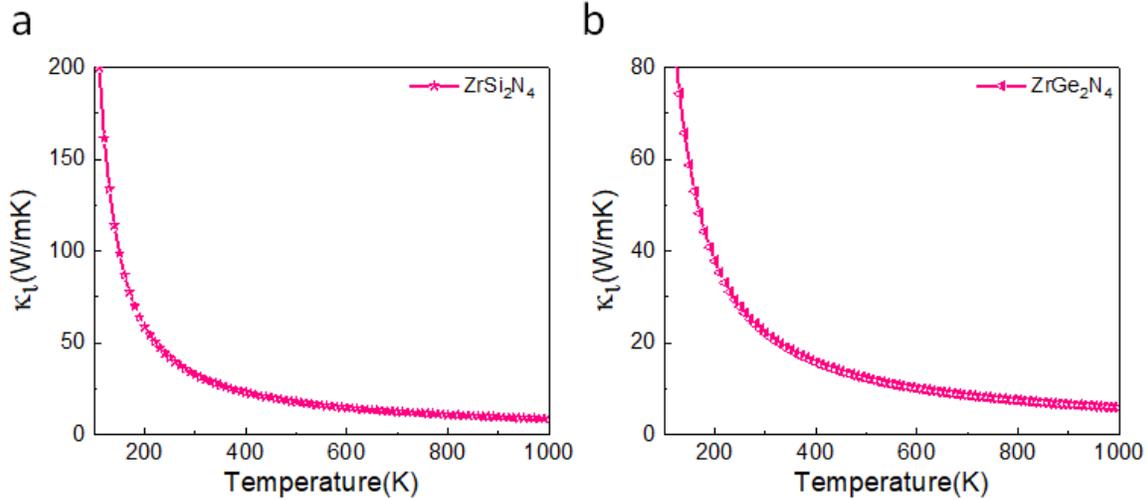

**Figure 6:** Graph illustrating the variation of $\kappa_l$ as a function of the temperature (K) for monolayer (a) $ZrSi_2N_4$ and (b) $ZrGe_2N_4$.

Figure 7 displays the relationship between the frequency and the phonon lifetime (τ) as well as the group velocity ($G_v$) for both monolayers. Figure 7a shows that the $ZrSi_2N_4$ has a significantly higher maximum phonon lifetime compared to the $ZrGe_2N_4$ monolayer shown in Figure 7b. However, the maximum phonon lifetime of $ZrGe_2N_4$ is low when compared to popular transition metal dichalcogenides (TMDCs) such as $WS_2$ and $MoS_2$ [34,35]. The $ZrSi_2N_4$ monolayer demonstrated a peak group velocity ($G_v$) of approximately 11.2 km/s (Figure 7c), primarily attributed to the longitudinal acoustic mode. Similarly, the $ZrGe_2N_4$ monolayer exhibited a peak $G_v$ value of approximately 10.9 km/s (Figure 7d) attributed to the longitudinal acoustic mode. Since the $\kappa_l$ value is directly related to both $G_v$ and τ, it is straightforward to explain the discrepancy in $\kappa_l$ values between the monolayers. Although the $G_v$ values are almost the same for both monolayers, the τ value of the $ZrSi_2N_4$ monolayer is significantly higher than that of $ZrGe_2N_4$. Consequently, the $\kappa_l$ value is larger in $ZrSi_2N_4$. The phonon band structure of $ZrSi_2N_4$ clearly exhibits distinct separation between the acoustic and optical phonon bands. Consequently, the likelihood of acoustic and optical phonon scattering is minimal, which is seen in the significant $\tau$ value and similarly high $\kappa_l$ value observed in the ZrSi$_2$N$_4$ monolayer. The presence of the heavier Ge atom in ZrGe$_2$N$_4$ resulted in a decrease in the total phonon frequencies when compared to ZrSi$_2$N$_4$. In ZrGe$_2$N$_4$, there is a simultaneous presence of low-lying optical phonons and acoustic phonons, with some degree of overlap. The increased scattering of acoustic phonons with optical phonons is seen in the low values for $\tau$ and the related low $\kappa_l$ value for the ZrGe$_2$N$_4$ monolayer [36].

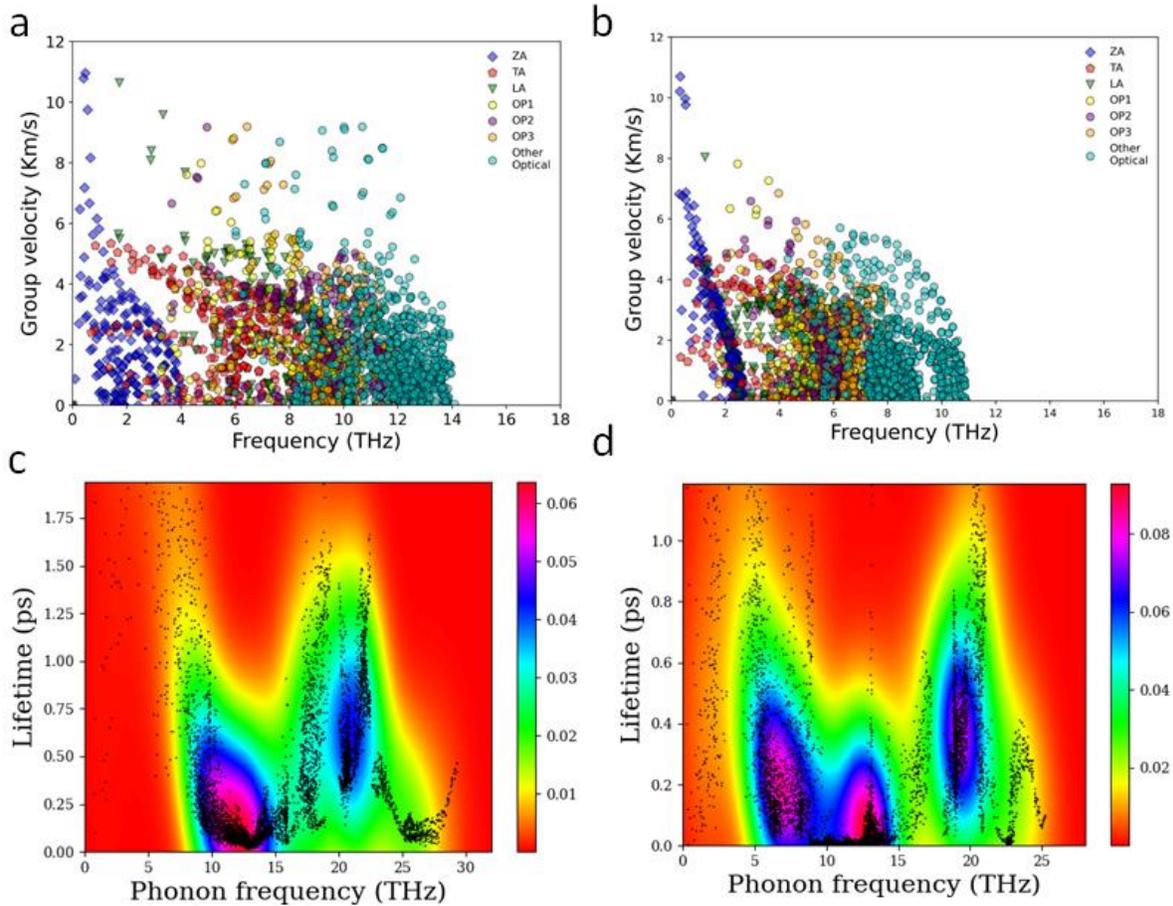

**Figure 7:** The frequency dependence of group velocity ($G_v$) for different acoustic and optical modes in (a) ZrSi$_2$N$_4$ and (b) ZrGe$_2$N$_4$ monolayers was examined. Frequency dependence of phonon lifetime (c) in ZrSi$_2$N$_4$ and (d) in ZrGe$_2$N$_4$ monolayers.

**Thermoelectric figure of merit (*ZT*)**

The $ZT$ is a measure of a material's efficiency with respect to the thermoelectric effect and its quality. In order to obtain a high $ZT$ value, the values of $\sigma$ and $S$ must be large, while the value of $k$ ($\kappa_e + \kappa_l$) must be significantly low, as shown in equation 8. The $ZT$ is characterized as

$$ZT = \frac{S^2 \sigma T}{\kappa_{el} + \kappa_l}, \qquad (8)$$

where $S, T, \sigma, k,$ and $k_l$ are identical to what was previously mentioned. Figure 8 (a) and (b) show, respectively, the $ZT$ product at 300 K, 600 K, and 900 K for the $ZrSi_2N_4$ and $ZrGe_2N_4$ monolayers, as a function of μ. The $ZrGe_2N_4$ monolayer exhibited the best $ZT$ value among them, with p-type of 0.90 and n-type of 0.83 at 900 K. In contrast, at 900 K, the $ZrSi_2N_4$ monolayer exhibited a p-type $ZT$ value of 0.89 and an n-type $ZT$ value of 0.82. With p-type doping, the $ZT$ values of both structures were higher than with n-type doping. According to Table 4, p-type doping was more effective than n-type doping for both structures because the $ZT$ value was higher for p-type doping. Although the $S$ of $ZrSi_2N_4$ is more than that of $ZrGe_2N_4$, the $\sigma$ of $ZrGe_2N_4$ is greater than that of $ZrSi_2N_4$, leading to a comparable power factor. However, the $\kappa_l$ played the essential role in transforming the $ZrGe_2N_4$ monolayer into a superior material compared to $ZrSi_2N_4$ towards thermoelectric figure of merit. Table 3 shows the maximum $ZT$ values for both monolayers at various temperatures.

**Table 3:** Calculated $ZT$ at different temperatures for monolayer of $ZrSi_2N_4$ and $ZrGe_2N_4$.

| Temperature | Thermoelectric figure of merit $ZT$ | | | |
| --- | --- | --- | --- | --- |
| K | $ZrSi_2N_4$ | | $ZrGe_2N_4$ | |
| | p | n | p | n |
| 300 | 0.54 | 0.36 | 0.69 | 0.36 |
| 600 | 0.82 | 0.71 | 0.85 | 0.72 |
| 900 | 0.89 | 0.82 | 0.90 | 0.83 |

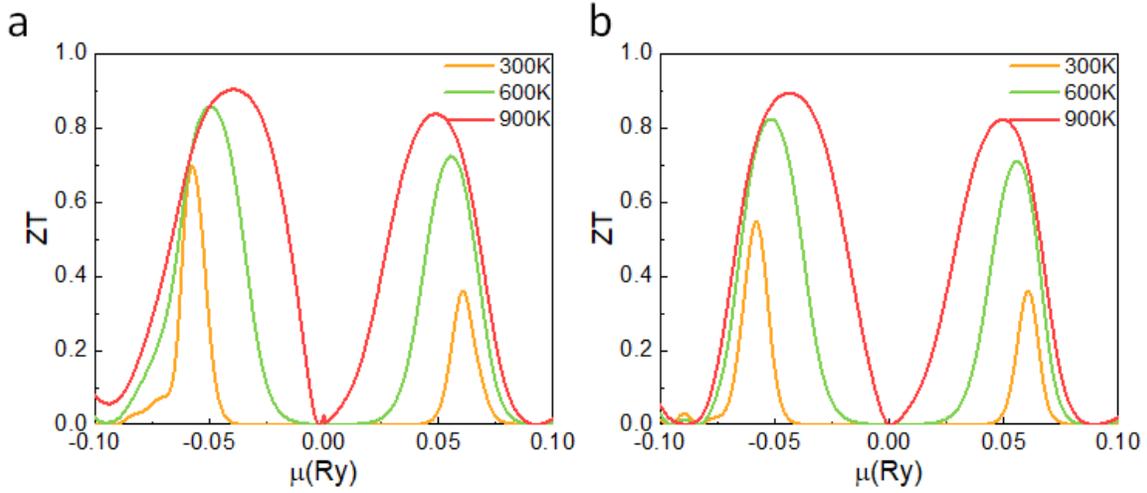

**Figure 8:** Plot of *ZT* at different temperatures (*T*) with respect to chemical potential (μ) for the monolayer of (a) ZrSi$_2$N$_4$ and (b) ZrGe$_2$N$_4$.

## 4. Conclusions

Using first-principles calculations, we determined the electrical, optical, and thermoelectric properties. The dynamical stability of both the monolayers was demonstrated by the fact that phonon dispersion curves did not contain any imaginary frequencies. Thermoelectric properties of ZrGe$_2$N$_4$, including its *ZT* and $\kappa_l$, are exceptional. In thermoelectric applications, however, ZrSi$_2$N$_4$ does not perform very well. At each of the three temperatures, the ZrGe$_2$N$_4$ produced excellent *ZT* products. The *ZT* product for the p-type was 0.69 at 300 K, 0.85 at 600 K, and 0.90 at 900 K. So, the ZrGe$_2$N$_4$ monolayer with p-type doping can be a potential element for thermoelectric devices. Therefore, ZrGe$_2$N$_4$ could be a candidate for next-generation thermoelectric devices that convert waste heat into energy at low-temperature as well as high-temperature thermoelectric devices to generate electricity from waste heat.

**Acknowledgments:**

We are thankful to the Department of Science and Technology (DST), India, for supporting us through the INSPIRE program and the Ministry of Human Resource and Development (MHRD). We are also grateful to the Indian Institute of Technology Jodhpur for providing the infrastructure to carry out the research.